\documentclass{article}
\usepackage{spconf,amsmath,graphicx}
\usepackage{amsfonts}
\usepackage{bm}
\usepackage{url}
\usepackage{multirow}
\usepackage{siunitx}
\usepackage{hyperref}
\newcolumntype{P}[1]{>{\centering\arraybackslash}p{#1}}

\title{Denoising-and-Dereverberation Hierarchical Neural Vocoder for Robust Waveform Generation}
\name{Yang Ai$^1$, Haoyu Li$^2$, Xin Wang$^2$, Junichi Yamagishi$^2$, Zhenhua Ling$^1$}
\address{
  $^1$University of Science and Technology of China, P.R.China,
  $^2$National Institute of Informatics, Japan}
\begin{document}
\ninept
\maketitle
\begin{abstract}
This paper presents a denoising and dereverberation hierarchical neural vocoder (DNR-HiNet) to convert noisy and reverberant acoustic features into a clean speech waveform.
We implement it mainly by modifying the amplitude spectrum predictor (ASP) in the original HiNet vocoder.
This modified denoising and dereverberation ASP (DNR-ASP) can predict clean log amplitude spectra (LAS) from input degraded acoustic features.
To achieve this, the DNR-ASP first predicts the noisy and reverberant LAS, noise LAS related to the noise information, and room impulse response related to the reverberation information then performs initial denoising and dereverberation. The initial processed LAS are then enhanced by another neural network as the final clean LAS.
To further improve the quality of the generated clean LAS, we also introduce a bandwidth extension model and frequency resolution extension model in the DNR-ASP.
The experimental results indicate that the DNR-HiNet vocoder was able to generate a denoised and dereverberated waveform given noisy and reverberant acoustic features and outperformed the original HiNet vocoder and a few other neural vocoders.
We also applied the DNR-HiNet vocoder to speech enhancement tasks, and its performance was competitive with several advanced speech enhancement methods.

\end{abstract}
%
\vspace{-2mm}
\section{Introduction}
\label{sec:intro}
\vspace{-0.5mm}

Neural vocoders, which reconstruct speech waveforms from acoustic features, are key components for text-to-speech (TTS) synthesis \cite{shen2018natural} and voice conversion (VC) systems \cite{liu2018wavenet}.
Typical neural vocoders use auto-regressive \cite{tamamori2017speaker,ai2018samplernn,lorenzo2018robust}, knowledge-distilling-based \cite{oord2017parallel,ping2018clarinet}, or flow-based neural networks \cite{prenger2018waveglow}. There are also non-autoregressive neural source-filter (NSF) \cite{wangNSFall} and HiNet vocoders \cite{ai2020neural,ai2020knowledge} that combine neural networks with signal processing algorithms.
The neural vocoders are trained to reconstruct a speech waveform $\bm{O}$ given acoustic features $\bm{C}$. In the generation stage, they produce a waveform $\widehat{\bm{O}}$ with $\widehat{\bm{C}}$ predicted from either the input text in a TTS system or the source speaker's speech in a VC system.
For both TTS and VC, neural vocoders require a high-quality, clean, and dry ${\bm{O}}$ and the corresponding $\bm{C}$.

However, waveforms captured in real-life scenarios are usually degraded by noise and reverberation.
To alleviate the impact of noise and reverberation on the input source speaker's speech waveform to a VC system, for example, we may first apply speech enhancement (SE) methods to the input waveform for denoising and dereverberation. We can then extract acoustic features from the enhanced signals, convert them towards the target speaker, and generate a waveform using neural vocoders trained on clean data.

Many deep-learning-based SE methods have been proposed to recover clean speech from degraded speech.
Earlier studies mainly focused on mapping-based and masking-based SE methods.
Mapping-based SE methods \cite{valentini2018speech,ai2019dnn,xu2014regression,kim2020t} map the spectral representations of degraded speech to those of clean speech without enhancing the phase spectra.
Masking-based SE methods predict the time-frequency (T-F) masks between the degraded and clean speech, and some of the latest methods, such as cIRM \cite{williamson2015complex} and RSM \cite{liu2019supervised}, can enhance the amplitude and phase spectra simultaneously.
Subsequently, some researchers adopted advanced deep-learning models to directly enhance degraded speech in the time domain \cite{su2019perceptually,rethage2018wavenet,pascual2017segan,macartney2018improved}.
In previous studies \cite{su2019perceptually,rethage2018wavenet}, neural waveform models inspired by WaveNet \cite{oord2016wavenet} were used for SE.
Such neural waveform enhancement models directly convert the degraded speech waveform $\bm{O}'$ into a clean speech waveform $\widehat{\bm{O}}$.
Obviously, there are similarities and common components between these models and neural vocoders. This inspired us to design a unified neural vocoder that can jointly perform denoising and dereverberation on degraded acoustic features and generate a clean speech waveform.
More specifically, this new vocoder is expected to generate a clean speech waveform $\widehat{\bm{O}}$ from noisy and reverberant conditional acoustic features $\bm{C}'$ (without using the noisy and reverberant speech waveform $\bm{O}'$).
This task is different from both pure neural vocoding and SE tasks, but such a joint model is convenient because it can perform these two tasks simultaneously and avoid the pipeline process that might propagate errors.

With the motivation above, we propose a denoising and dereverberation HiNet (DNR-HiNet) vocoder.
Similar to the original HiNet vocoder, the DNR-HiNet vocoder uses an amplitude spectrum predictor (ASP) and phase spectrum predictor (PSP) to predict amplitude and phase spectra, respectively, and it reconstructs the waveform through short-time Fourier synthesis (STFS).
However, the DNR-HiNet vocoder combines the original ASP with a channel encoder, noise encoder, reverberation encoder, and post-denoising-dereverberation module so that the modified ASP module can predict clean amplitude spectra from noisy and reverberant acoustic features.
To boost the generated audio quality, the DNR-HiNet further adds a bandwidth extension (BWE) model and frequency resolution extension (FRE) model to the modified ASP.
Experiments confirmed that the DNR-HiNet vocoder outperformed baseline neural vocoders. It also showed competitive performance with several SE methods on the SE task.

This paper is organized as follows: In Section \ref{sec: Previous works}, we briefly review the original HiNet vocoder. In Section \ref{sec: Proposed methods}, we give details on our proposed DNR-HiNet vocoder. In Section \ref{sec: Experiments}, we present our experiments and results. Finally, we give conclusions in Section \ref{sec: Conclusion}.

\vspace{-0.5mm}
\section{Brief overview of original HiNet vocoder}
\label{sec: Previous works}

\vspace{-1mm}
\subsection{Original HiNet vocoder}
\label{subsec: Original HiNet vocoder}
\vspace{-1mm}

Given input acoustic features, the original HiNet vocoder uses an ASP and a PSP to predict frame-level log amplitude spectra (LAS) and phase spectra, respectively \cite{ai2020knowledge}. It then reconstructs the waveform from the predicted amplitude and phase spectra through STFS.

\begin{figure}[t]
    \centering
    \includegraphics[width=1\columnwidth]{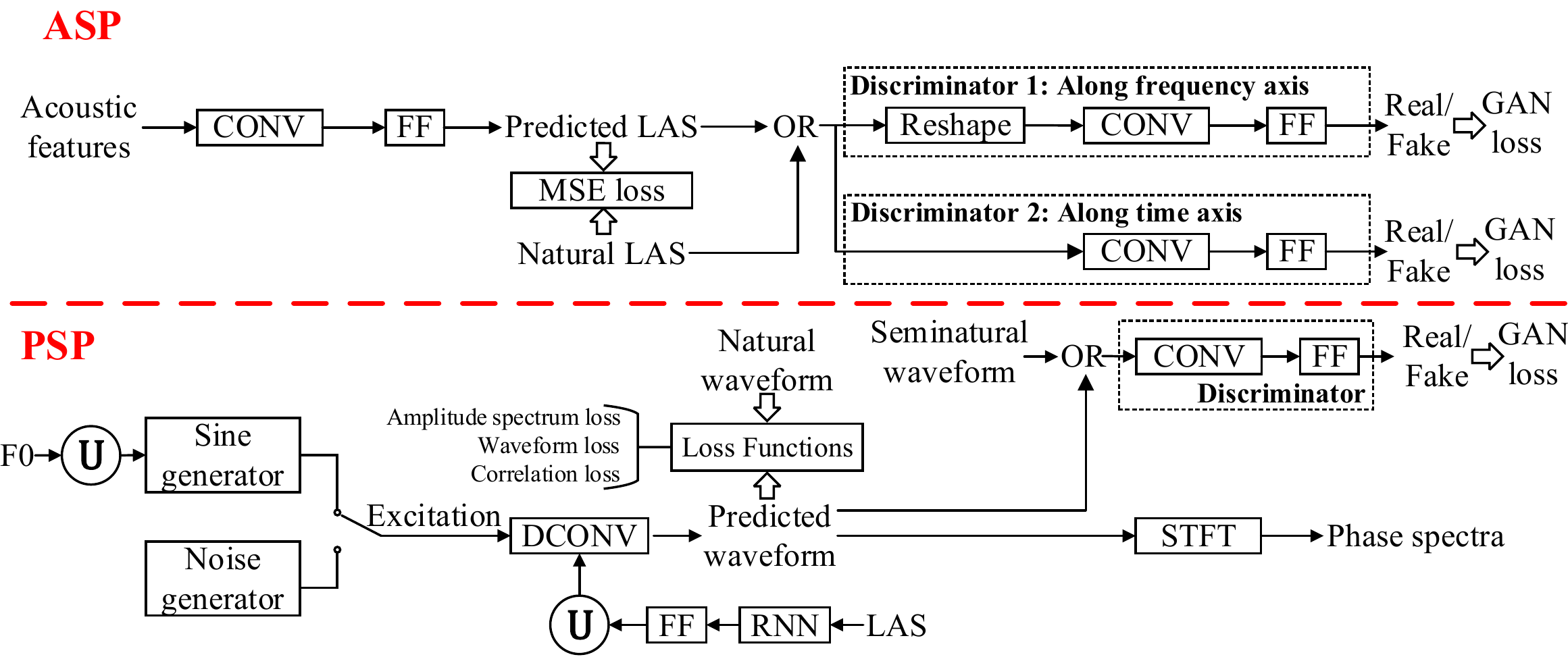}
    \caption{Flowchart of ASP and PSP in original HiNet vocoder. {}{FF}, {}{RNN}, {}{CONV} and {}{DCONV} denote feed-forward, unidirectional GRU-based recurrent layers, convolutional layers and dilated convolutional layers, respectively. \textbf{U} denotes upsampling operation.}
    \label{fig: ASP}
    \vspace{-5mm}
\end{figure}

The structures of the ASP and PSP are plotted in Figure \ref{fig: ASP}.
The ASP is based on a generative adversarial network (GAN), where both the generator and the two discriminators include multiple convolutional layers.
One of the discriminators conducts convolution along the frequency axis of the input LAS, while the other operates along the time axis.
The PSP is also based on a GAN. Its generator is similar to the NSF vocoder \cite{wangNSFall} and converts the LAS and F0 into a preliminary waveform for phase extraction.
A discriminator discriminates the preliminary waveform from a seminatural waveform reconstructed using the natural phase spectra and amplitude spectra of a predicted waveform by a well-trained PSP without GANs.

\vspace{-1mm}
\subsection{HiNet vocoder for reverberation modeling}
\label{subsec: HiNet vocoder for reverberation modeling}
\vspace{-1mm}

Recently, a HiNet vocoder with a trainable reverberation module was proposed for better reverberation modeling \cite{ai2020reverberation}.
This module produces a reverberant waveform by convolving the waveform from the generator in the PSP with a room impulse response (RIR). The phase spectra from the reverberant waveform are then used for STFS.

The RIR is predicted by another trainable neural network from the input LAS, and this RIR predictor is jointly trained with the PSP by minimizing the distances between the generated and natural reverberant waveforms.
Experimental results confirmed that the RIR could be estimated and that reverberation module was helpful for modeling the reverberation effect.

\vspace{-0.5mm}
\section{Proposed DNR-HiNet vocoder}
\label{sec: Proposed methods}

\begin{figure*}[t]
    \centering
    \includegraphics[width=2\columnwidth]{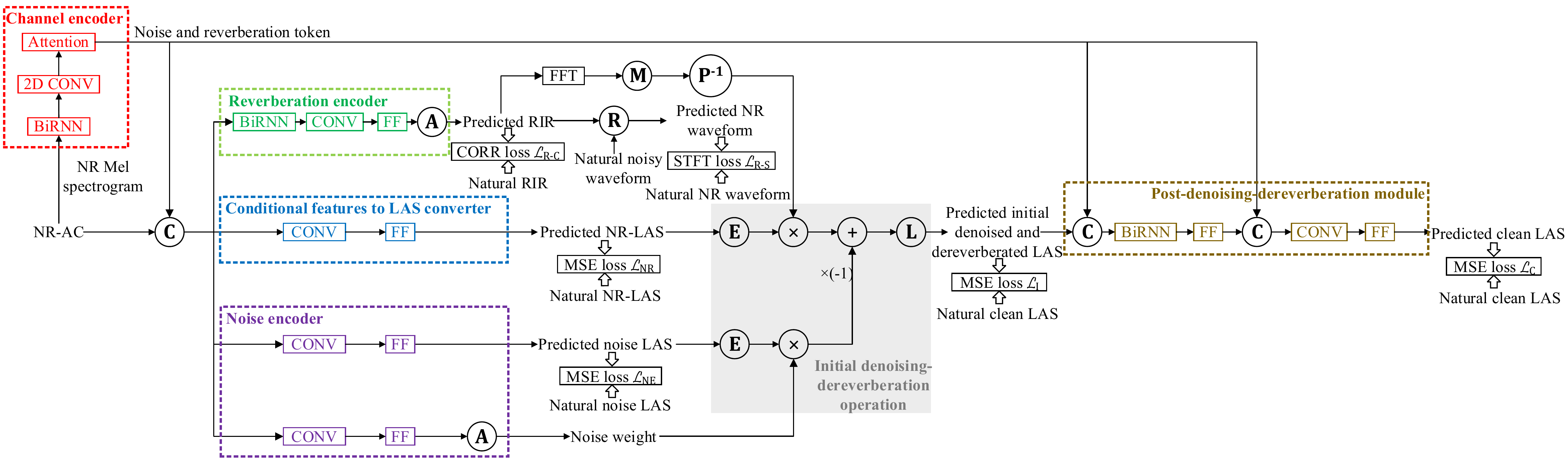}
    \vspace{-3mm}
    \caption{Flowchart of DNR-ASP. Discriminators are not plotted for simplicity. ``NR'' and ``AC'' stand for ``noisy and reverberant'' and ``acoustic features'', respectively. LAS denotes log amplitude spectral sequence. {}{FF}, {}{CONV}, and {}{BiRNN} denote fully connected feed-forward, convolutional, and bidirectional GRU-based recurrent layers, respectively.
    \textbf{A}, \textbf{C}, \textbf{E}, \textbf{L}, \textbf{M}, \textbf{P$^{-1}$}, and \textbf{R} denote temporal averaging, concatenation, natural exponential function, logarithm, magnitude, reciprocal and convolution, respectively. Symbols $\times$ and $+$ denote element-wise multiplication and addition, respectively.}
    \label{fig: ASP_DNR}
    \vspace{-3mm}
\end{figure*}

\subsection{DNR-HiNet vocoder}
\label{subsec: DNR-HiNet vocoder}
\vspace{-1mm}
Compared with the original HiNet vocoder, our DNR-HiNet vocoder uses a modified ASP, which we refer to as the DNR-ASP.
The DNR-ASP conducts two stages of denoising and dereverberation to convert the input noisy and reverberant acoustic features into clean LAS, after which the PSP predicts the phase spectra from the estimated clean LAS.
In the first stage, the DNR-ASP estimates the factorized noise and reverberation information from the input and produces an initial clean LAS using neural networks and other deterministic operations.
In the second stage, the DNR-ASP uses another neural-network module to enhance the initial clean LAS.
The DNR-ASP requires clean, noisy, and noisy and reverberant waveforms during training. Therefore, we use artificially generated noisy and reverberant speech for training. However, the trained model can be used for real noisy and reverberant speech.

\vspace{-1mm}
\subsubsection{Structure of DNR-ASP}
\label{subsubsec: Structures of DNR-ASP}
\vspace{-1mm}

The structure of the DNR-ASP is illustrated in Figure \ref{fig: ASP_DNR}. It includes neural-network-based modules (color boxes), deterministic operations (black circles), and loss functions (black square boxes). All the neural-network-based modules are jointly trained using the loss functions and GAN-based criteria (not shown in the figure for simplicity). The details are explained in the following paragraphs.

\noindent
{}{\textbf{Channel encoder}}:
This module is inspired by a previous study \cite{li2020noise}, and used to encode the noise and reverberation information that is useful for other modules.
It uses the 80-dimensional noisy and reverberant mel-spectrogram as input and generates a noisy and reverberant token.
The token is expected to be relevant with the noise and reverberation information extracted from the mel-spectrogram.
The aim is to distinguish different types of noise and reverberation and generalize with unseen types in the test set.
This module consists of a bidirectional GRU-based recurrent layer with 80 nodes (40 for forward and 40 for backward), 5 2-D convolutional layers with 32, 64, 64, 128, and 256 channels respectively, and a multi-head attention model \cite{vaswani2017attention}.
In the convolutional layers, the filter size is 5(time axis)$\times$5(frequency axis) for each layer, and the stride is 1$\times$2 for the first 4 layers and 1$\times$5 for the last layer.
In the attention model, referring to the above study \cite{li2020noise}, the 256-dimensional output of the last convolutional layer is served as the $\emph{query}$ and 16 trainable 256-dimensional noise and reverberation templates are served as the $\emph{key}$ and $\emph{value}$.
The number of attention heads is set to 8.
The attention model finally generates a 256-dimensional noise and reverberation token which is the weighted sum of the noise and reverberation templates.
The resulting token is used as the additional input of other modules.

\vspace{0.5mm}
\noindent
{}{\textbf{Conditional features to LAS converter}}:
This module is the same as the original ASP model and used to predict the noisy and reverberant LAS $\hat{\bm{L}}^{NR}=[\hat{\bm{L}}_1^{NR},\dots,\hat{\bm{L}}_N^{NR}]^\top$ (where $\hat{\bm{L}}_n^{NR}=[\hat{L}_{n,1}^{NR},\dots,\hat{L}_{n,K}^{NR}]^\top$)\footnote{If there are no special instructions, $\bm{X}=[\bm{X}_1,\dots,\bm{X}_N]^\top$, where $\bm{X}_n=[X_{n,1},\dots,X_{n,K}]^\top$.} from the input noisy and reverberant acoustic features and token.
Here, $N$ and $K=\frac{FN}{2}+1$ represent the total number of frames and frequency bins, respectively, and $n\in\{1,\dots,N\}$ is the frame index; $FN$ is the FFT point number.
This module consists of 2 convolutional layers with 2048 nodes per layer (filter width=7) and a $K$-
dimensional feed-forward (FF) linear output layer.
The loss function is the mean square error (MSE) between $\hat{\bm{L}}^{NR}$ and the natural noisy and reverberant LAS $\bm{L}^{NR}$, i.e.,
\vspace{-3mm}
\begin{equation}
\label{equ: L_NR}
\mathcal L_{NR}=\dfrac{1}{NK}\sum_{n=1}^{N}\sum_{k=1}^{K}(L_{n,k}^{NR}-\hat{L}_{n,k}^{NR})^2.
\end{equation}
\vspace{-3mm}

\noindent
{}{\textbf{Noise encoder}}:
Given the noisy and reverberant acoustic features and token, this module explicitly estimates the noise information for initial denoising and is divided into two sub-modules.
The first sub-module estimates noise LAS $\hat{\bm{L}}^{NE}$ using 2 convolutional layers (1024 nodes per layer, filter width=7) and a $K$-dimensional FF linear output layer.
The loss function is the MSE between $\hat{\bm{L}}^{NE}$ and natural noise LAS $\bm{L}^{NE}$, i.e.,
\vspace{-2mm}
\begin{equation}
\label{equ: L_NE}
\mathcal L_{NE}=\dfrac{1}{NK}\sum_{n=1}^{N}\sum_{k=1}^{K}(L_{n,k}^{NE}-\hat{L}_{n,k}^{NE})^2.
\vspace{-3mm}
\end{equation}
The second sub-module predicts the weight of the noise amplitude spectra $\alpha$ using a convolutional layer (256 nodes, filter width=5), 1-dimensional FF linear layer, and temporal average pooling layer (i.e., average from $n=1$ to $n=N$).

\vspace{0.5mm}
\noindent
{}{\textbf{Reverberation encoder}}:
Inspired by the HiNet vocoder with reverberation modeling mentioned in Section~\ref{subsec: HiNet vocoder for reverberation modeling}, this module estimates an RIR vector $\hat{\bm{r}}=[\hat{r}_1,\dots,\hat{r}_{FN}]^\top$ given the input noisy and reverberant acoustic features and token.
This module consists of a bidirectional GRU-based recurrent layer with 256 nodes (128 for forward and 128 for backward), convolutional layer with 1024 nodes (filter width=9), $FN$-dimensional FF linear layer, and temporal average pooling layer.
The module has two specific loss functions.
The first one is defined as the negative correlation coefficient between $\hat{\bm{r}}$ and natural RIR $\bm{r}=[r_1,\dots,r_{FN}]^\top$, i.e.,
\begin{equation}
\label{equ: L_R-C}
\mathcal L_{R-C}=-\dfrac{\mathbb{E}[(\bm{r}-\mathbb{E}(\bm{r}))(\hat{\bm{r}}-\mathbb{E}(\hat{\bm{r}}))]}{\sqrt{\mathbb{V}(\bm{r})\mathbb{V}(\hat{\bm{r}})}},
\end{equation}
where $\mathbb{E}(\cdot)$ and $\mathbb{V}(\cdot)$ calculate mean and variance, respectively.
Minimizing $L_{R-C}$ is expected to increase the similarity between $\hat{\bm{r}}$ and $\bm{r}$.
To further improve $\hat{\bm{r}}$, the second metric $L_{R-S}$ is defined as the multi-scale STFT loss \cite{ai2020reverberation} between the natural noisy and reverberant waveform and a predicted one. This predicted version is acquired by convolving the natural noisy waveform with $\hat{\bm{r}}$ in the frequency domain (i.e., \textbf{R} in Figure \ref{fig: ASP_DNR}).

\vspace{0.5mm}
\noindent
{}{\textbf{Initial denoising-dereverberation operation}}:
We assume that a noisy and reverberant waveform is obtained from a clean waveform by convolving it with RIR and adding the noise in the time domain.
For denoising and dereverberation, given the noise and reverberation information estimated by the noise and reverberation modules, we initially remove the noise and reverberation effect from $\hat{\bm{L}}^{NR}$ through operations in the spectral domain:
\vspace{-1mm}
\begin{equation}
\label{equ: I}
\tilde{L}_{n,k}^{C}=\log [\exp(\hat{L}_{n,k}^{NR}) \cdot \dfrac{1}{\hat{R}_k}-\alpha \cdot \exp(\hat{L}_{n,k}^{NE})],
\end{equation}
where $\hat{R}_k$ is the magnitude of the $FN$-point FFT of $\hat{\bm{r}}$ at the $k$-th frequency bin, $k\in\{1,\dots,K\}$, and $n\in\{1,\dots,N\}$.
This process is highlighted in the grey region of Figure \ref{fig: ASP_DNR}.
We also define the MSE loss between the initially processed (denoised and dereverberated) LAS $\tilde{\bm{L}}^{C}$ and natural clean LAS $\bm{L}^{C}$, i.e.,
\vspace{-2mm}
\begin{equation}
\label{equ: L_C_I}
\mathcal L_{I}=\dfrac{1}{NK}\sum_{n=1}^{N}\sum_{k=1}^{K}(L_{n,k}^{C}-\tilde{L}_{n,k}^{C})^2.
\end{equation}
\vspace{-2mm}

\noindent
{}{\textbf{Post-denoising-dereverberation module}}:
Because the linear operations in Eq.~(\ref{equ: I}) may not remove all the reverberation and noise, we add another neural-network-based module to further enhance $\tilde{\bm{L}}^{C}$ and produce the final clean LAS $\hat{\bm{L}}^{C}$ frame by frame.
This module consists of a bidirectional GRU-based recurrent layer with 1024 nodes (512 for forward and 512 for backward), a $K$-dimensional FF linear layer, 2 convolutional layers each with 2048 nodes (filter width=7), and a $K$-dimensional FF linear output layer.
The input of the module includes $\tilde{\bm{L}}^{C}$ and the token.
The loss function is defined as the MSE between $\hat{\bm{L}}^{C}$ and $\bm{L}^{C}$, i.e.,
\vspace{-2mm}
\begin{equation}
\label{equ: L_C}
\mathcal L_{C}=\dfrac{1}{NK}\sum_{n=1}^{N}\sum_{k=1}^{K}(L_{n,k}^{C}-\hat{L}_{n,k}^{C})^2.
\end{equation}

\vspace{-3mm}
\subsubsection{Training criteria of DNR-ASP}
\label{subsubsec: Training criteria of DNR-ASP}
\vspace{-1mm}

Following a previous study \cite{ai2020reverberation}, a Wasserstein GAN \cite{gulrajani2017improved} loss with gradient penalty \cite{mescheder2018training} is used to train the DNR-ASP.
Two discriminators that share the same structures as those in Figure \ref{fig: ASP} are adopted.
The training process is divided into three steps.
First, use $\mathcal L_{NR}+\mathcal L_{NE}+\mathcal L_{R-C}+\mathcal L_{R-S}+\mathcal L_{I}+\mathcal L_{C}$ to jointly update all the trainable modules in Figure~\ref{fig: ASP_DNR};
second, use the discriminator loss $\mathcal L_{D}$ to train the discriminators. Finally, use $\mathcal L_{G-adv}+\lambda \cdot \mathcal L_{C}$ and $\mathcal L_{D}$ to train the post-denoising-dereverberation module and discriminators alternately, where $\mathcal L_{G-adv}$ is the adversarial loss from the discriminators and $\lambda$ is a hyperparameter ($\lambda=500$ in the experiments).

\vspace{-1mm}
\subsubsection{Structure and training criteria of PSP}
\label{subsubsec: Structures and training criteria of PSP}
\vspace{-1mm}
Regarding the PSP, we use the original PSP  \cite{ai2020neural,ai2020knowledge} with two modifications.
First, we add MSE on LAS to the original loss functions which include MSE on amplitude spectra, waveform loss, and correlation loss.
Second, the real input of the discriminator is the natural waveform instead of the seminatural one.
While being trained using the clean data, the PSP at the generation stage takes noisy and reverberant F0\footnote{We confirmed that the degradation due to using noisy and reverberant F0 as the PSP's input rather than clean F0 was minor.} and clean LAS predicted using the DNR-ASP as input.

\vspace{-1mm}
\subsection{DNR-HiNet vocoder with BWE and FRE models}
\label{subsec: DNR-HiNet vocoder with BWE and FRE models}
\vspace{-1mm}

\begin{figure}[t]
    \centering
    \includegraphics[width=1\columnwidth]{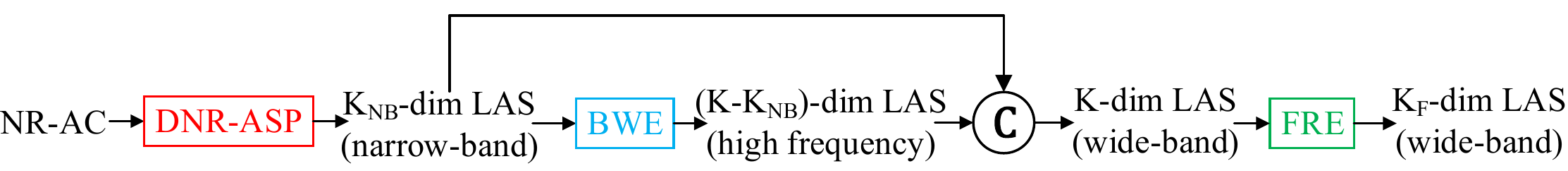}
    \caption{Flowchart of DNR-ASP with BWE and FRE models. Symbols and abbreviations are same as those in Figure \ref{fig: ASP_DNR}.}
    \label{fig: ASP_BWE_FRE}
    \vspace{-5mm}
\end{figure}

The middle and high-frequency bands of speech signals contain aperiodic components, and the denoising process may wrongly remove such natural stochastic components from the speech signals.
Therefore, we first train the DNR-ASP using narrow-band LAS then train a BWE afterward to expand the frequency ranges of the enhanced narrow-band LAS from the DNR-ASP. This is expected to better reproduce the aperiodic characteristics of speech signals.
In addition, speech synthesis and VC tasks normally use finer frequency resolution than typical SE tasks. Therefore, we use an additional FRE model to increase the frequency resolution after the enhancement.
The entire process is illustrated in Figure~\ref{fig: ASP_BWE_FRE}.

We use $\hat{\bm{L}}^{C-NB}=[\hat{\bm{L}}_1^{C-NB},\dots,\hat{\bm{L}}_N^{C-NB}]^\top$ to denote the narrow-band LAS, where $\hat{\bm{L}}_n^{C-NB}=[\hat{L}_{n,1}^{C},\dots,\hat{L}_{n,K_{NB}}^{C}]^\top$ and $K_{NB}<K$.
The BWE model first predicts the high-frequency LAS $\hat{\bm{L}}^{C-HF}=[\hat{\bm{L}}_1^{C-HF},\dots,\hat{\bm{L}}_N^{C-HF}]^\top$ (where $\hat{\bm{L}}_n^{C-HF}=[\hat{L}_{n,K_{NB}}^{C},\dots,\hat{L}_{n,K}^{C}]^\top$) from the input $\hat{\bm{L}}^{C-NB}$.
The wide-band LAS $\hat{\bm{L}}^{C}$ are then obtained by concatenating $\hat{\bm{L}}^{C-NB}$ and $\hat{\bm{L}}^{C-HF}$.
Finally, the FRE model extends the frequency resolution (i.e., interpolation) of $\hat{\bm{L}}^{C}$ to $\hat{\bm{L}}^{C-F}=[\hat{\bm{L}}_1^{C-F},\dots,\hat{\bm{L}}_N^{C-F}]^\top$, where $\hat{\bm{L}}_n^{C-F}=[\hat{L}_{n,1}^{C-F},\dots,\hat{L}_{n,K_{F}}^{C-F}]^\top$ and $K_F>K$.

Both BWE and FRE models include 2 bidirectional GRU layers each with 1024 nodes (512 for forward and 512 for backward), 2 convolutional layers each with 2048 nodes (filter width=9), and an FF linear output layer ($K$ nodes for BWE model and $K_F$ nodes for FRE model).
While both models are trained using the clean data, the GANs used in the DNR-ASP are applied to the BWE model but not to the FRE model.
When BWE and FRE models are used, the PSP is also trained using $K_F$-dimensional LAS as input.

\section{Experiments}
\label{sec: Experiments}

\vspace{-1mm}
\subsection{Data and feature configuration}
\label{subsec: Data and feature configuration}
\vspace{-1mm}

We prepared a noisy and reverberant speech database based on the clean speech corpus\footnote{\url{http://dx.doi.org/10.7488/ds/2117}} in \cite{valentini2018speech}.
From this corpus, we selected a subset that contains 11,572 clean waveforms from 28 speakers and randomly divided the subset into a training set (11,012 utterances) and validation set (560 utterances).
We then selected another subset as the test set, which included 824 utterances from 2 unseen speakers.

We created noisy and reverberant speech data by adding noise and convolving RIR to the clean data.
Regarding the training and validation sets, all ten noise types and four SNRs used in \cite{valentini2018speech} were adopted in our experiments.
The RIRs at two microphone positions were selected from five rooms: booth (210ms), office (574ms), and lecture (827ms) in AIR database \cite{jeub2009binaural}, medium size room (341ms) in MIRD database \cite{hadad2014multichannel}, and reflective room (523ms)\footnote{The number in ($\cdot$) represented the average reverberation time T60 on two microphone positions calculated using the tools provided in \cite{hadad2014multichannel}.} in MARDY  database \cite{wen2006evaluation}.
Regarding the test set, all the five noise types and four SNRs used in \cite{valentini2018speech} were used.
The RIRs at two microphone positions were selected from three rooms: small room (188ms) and large room (593ms) in MIRD database \cite{hadad2014multichannel}, and meeting room (300ms) in AIR database \cite{jeub2009binaural}.
Note that all the noise types, SNRs, and RIRs of the test set were unseen in the training and validation sets.
We used the same tools in \cite{valentini2018speech} to add the noise and reverberation effects.
We also saved the corresponding noise and noisy speech data (i.e., clean speech + noise) for training the DNR-ASP.

The original 48-kHz clean waveforms were down sampled to 24-kHz for the construction of the noisy and reverberant database and the experiments.
The input noisy and reverberant acoustic features included the 80-dimensional mel-spectrogram, F0 extracted using YAPPT \cite{kasi2002yet}, and a voiced/unvoiced flag.
For training the DNR-ASP, we extracted the target clean LAS, noise LAS, and noisy and reverberant LAS from the natural clean, noise, and noisy and reverberant waveforms, respectively, using 2048 FFT points (i.e., $FN=2048$ and $K=1025$).
For training the FRE model, we extracted the target clean LAS from the natural clean waveform using 8192 FFT points (i.e., $K_F=4097$).
All the above features were extracted with a frame shift and length of 12 and 50 ms, respectively.

\vspace{-1mm}
\subsection{Experimental models}
\label{subsec: Experimental models}
\vspace{-1mm}

We first compared the following vocoders\footnote{Examples of generated speech can be found at \url{http://home.ustc.edu.cn/~ay8067/DNR/demo.html}.}.

\noindent
{}{\textbf{Baseline-NSF}}: The harmonic-plus-noise NSF model \cite{wang2019} was included as a reference model.
The input was the noisy and reverberant acoustic features, and the training target was the clean waveform.
We used the Pytorch re-implementation of NSF \footnote{\url{https://github.com/nii-yamagishilab/project-NN-Pytorch-scripts}} and added two bi-directional LSTM recurrent layers with 128 nodes to process the input acoustic features at the frame level. The number of model parameters was around \num{9.1e5}.
We also trained the baseline NSF model using the noisy and reverberant speech as the output instead of the clean speech as the lower-bound model,
which we referred to as \textbf{Baseline-NSF$'$}.

\noindent
{}{\textbf{Baseline-HiNet}}: The baseline HiNet vocoder.
The input and output of the ASP were the noisy and reverberant acoustic features and clean LAS, respectively.
The number of model parameters was around \num{3.2e7} for the ASP and \num{7.2e6} for the PSP.
We also trained the corresponding lower-bound model \textbf{Baseline-HiNet$'$} using the noisy and reverberant speech as the output.

\noindent
{}{\textbf{DNR-HiNet}}: Our proposed DNR-HiNet vocoder.
The model configuration was introduced in Section \ref{subsec: DNR-HiNet vocoder}.
The DNR-ASP has \num{1.1e8} trainable parameters, and the PSP is identical to that in baseline HiNet.

\noindent
{}{\textbf{DNR-HiNet w/ BF}}: Our proposed DNR-HiNet vocoder with BWE and FRE models.
The model configuration was introduced in Section \ref{subsec: DNR-HiNet vocoder with BWE and FRE models}, where the narrow-band frequency was 4000 Hz (i.e., $K_{NB}=341$).
The number of model parameters was around \num{9.4e7} for the ASP, \num{6.8e7} for the BWE model, \num{7.2e7} for the FRE model, and \num{1.6e7} for the PSP.
Note that although the model size of {\textbf{DNR-HiNet}} and {\textbf{DNR-HiNet w/ BF}} is huge, most of the calculation is carried out at the frame level, which does not significantly degrade generation efficiency.

We then compared the proposed model with advanced SE methods.
The difference between vocoders and SE methods is that the former uses only noisy and reverberant acoustic features, which may be predicted from speech synthesis or voice conversion models. On the other hand, the SE methods can use the noisy and reverberant waveform or more detailed representations such as the phase extracted from the waveform as input. In other words, the SE methods can leverage more acoustic features from the input waveform.
The SE methods used in the experiments were as follows:

\noindent
{}{\textbf{cIRM}}: An SE method based on complex ratio masking (cIRM) \cite{williamson2015complex}.
We reimplemented it using the same model architecture with the FRE model used in \textbf{DNR-HiNet w/ BF}.
The input and output of this model were the noisy and reverberant LAS and cIRM, respectively.
The number of model parameters was around \num{6.8e7}.

\noindent
{}{\textbf{SEGAN}}: A waveform-to-waveform mapping SE method based on GANs \cite{pascual2017segan}.
We trained it using the official open source implementation\footnote{\url{https://github.com/santi-pdp/segan}}.
The number of model parameters was around \num{7.3e7}.

\noindent
{}{\textbf{WaveNet}}: A waveform-to-waveform mapping SE method based on denoising WaveNet \cite{su2019perceptually}.
We re-implemented it with the same model architecture and training criterion (L1 loss on log spectrogram).
The number of model parameters was around \num{8.4e6}.

\noindent
{}{\textbf{T-GSA}}: A real value masking-based SE method based on the Guassian-weighted Transformer structure \cite{kim2020t}, which consists of 10 Transformer blocks each having 2048 input and output nodes. The soft mask was first predicted from input noisy and reverberant amplitude spectra then multiplied with the input amplitude spectra to produce the enhanced amplitude spectra. The enhanced waveform is then synthesized from the enhanced amplitude spectra by STFS with input phase spectra.
The number of model parameters was around \num{3.2e7}.

\noindent
{}{\textbf{DNR-HiNet$^{*}$ w/ BF}}:
Same as \textbf{DNR-HiNet w/ BF} except that the phase spectra are extracted from the input noisy and reverberant waveform rather than being predicted using the PSP.
This is another way to use DNR-HiNet for the SE task in which the input noisy and reverberant waveform is known.
It is also for a fair comparison between the proposed DNR-HiNet and other SE methods that explicitly or implicitly use the phase in the noisy and reverberant waveform, i.e., \textbf{SEGAN}, \textbf{WaveNet}, and \textbf{T-GSA}.

\vspace{-1mm}
\subsection{Comparison among neural vocoders}
\label{subsec: Comparison among neural vocoders}
\vspace{-1mm}

We conducted this experiment to compare the performance of the neural vocoders mentioned in Section \ref{subsec: Experimental models} for the denoising and dereverberation task. The model input is noisy and reverberant acoustic features.

\vspace{-1mm}
\subsubsection{Objective evaluation}
\label{subsec: Objective evaluation}
\vspace{-1mm}

We first compared the performance of these neural vocoders using four objective metrics, including the short-time objective intelligibility (STOI) score \cite{taal2010short}, which reflects the speech intelligibility, and three composite measures (CSIG, CBAK and COVL) \cite{hu2006evaluation}, which predict the mean opinion score (MOS) on signal distortion, background noise intrusiveness, and overall effect, respectively. For all four metrics, a higher score indicates better performance.

\begin{table}
\vspace{-3mm}
\centering
    \caption{Objective evaluation results of different systems. Here, the numbers in the table represent the average value on test set.}
    \renewcommand\arraystretch{1.05}
    \resizebox{8.0cm}{2.2cm}{
    \begin{tabular}{l | c c c c}
        \hline
        \hline
         & STOI & CSIG & CBAK & COVL \\
         \hline
         Noisy and reverberant audio& 0.777 & 2.21 & 1.84 & 2.05\\
         \hline
         \textbf{Baseline-NSF$'$} & 0.740 & 1.91 & 1.59 & 1.70 \\
         \textbf{Baseline-NSF} & 0.763 & 2.99 & 1.98 & 2.37 \\
         \textbf{Baseline-HiNet$'$} & 0.746 & 2.18 & 1.76 & 1.99 \\
         \textbf{Baseline-HiNet} & 0.705 & 2.99 & 2.06 & 2.48 \\
         \textbf{DNR-HiNet} & 0.769 & \textbf{3.25} & 2.24 & 2.69 \\
         \textbf{DNR-HiNet w/ BF} & \textbf{0.783} & 3.24 & \textbf{2.29} & \textbf{2.75} \\
         \hline
         \textbf{cIRM} & 0.701 & 2.24 & 1.81 & 1.98 \\
         \textbf{SEGAN} & 0.659 & 1.76 & 1.26 & 1.55 \\
         \textbf{WaveNet} & 0.800 & 3.35 & 2.35 & 2.78 \\
         \textbf{T-GSA} & \textbf{0.818} & 3.32 & 2.43 & 2.87 \\
         \textbf{DNR-HiNet$^*$ w/ BF} & 0.803 & \textbf{3.38} & \textbf{2.44} & \textbf{2.92} \\
        \hline
        \hline
    \end{tabular}}
\label{tab_objective}
\vspace{-5mm}
\end{table}

\begin{figure}[t]
    \centering
    \includegraphics[width=1\columnwidth]{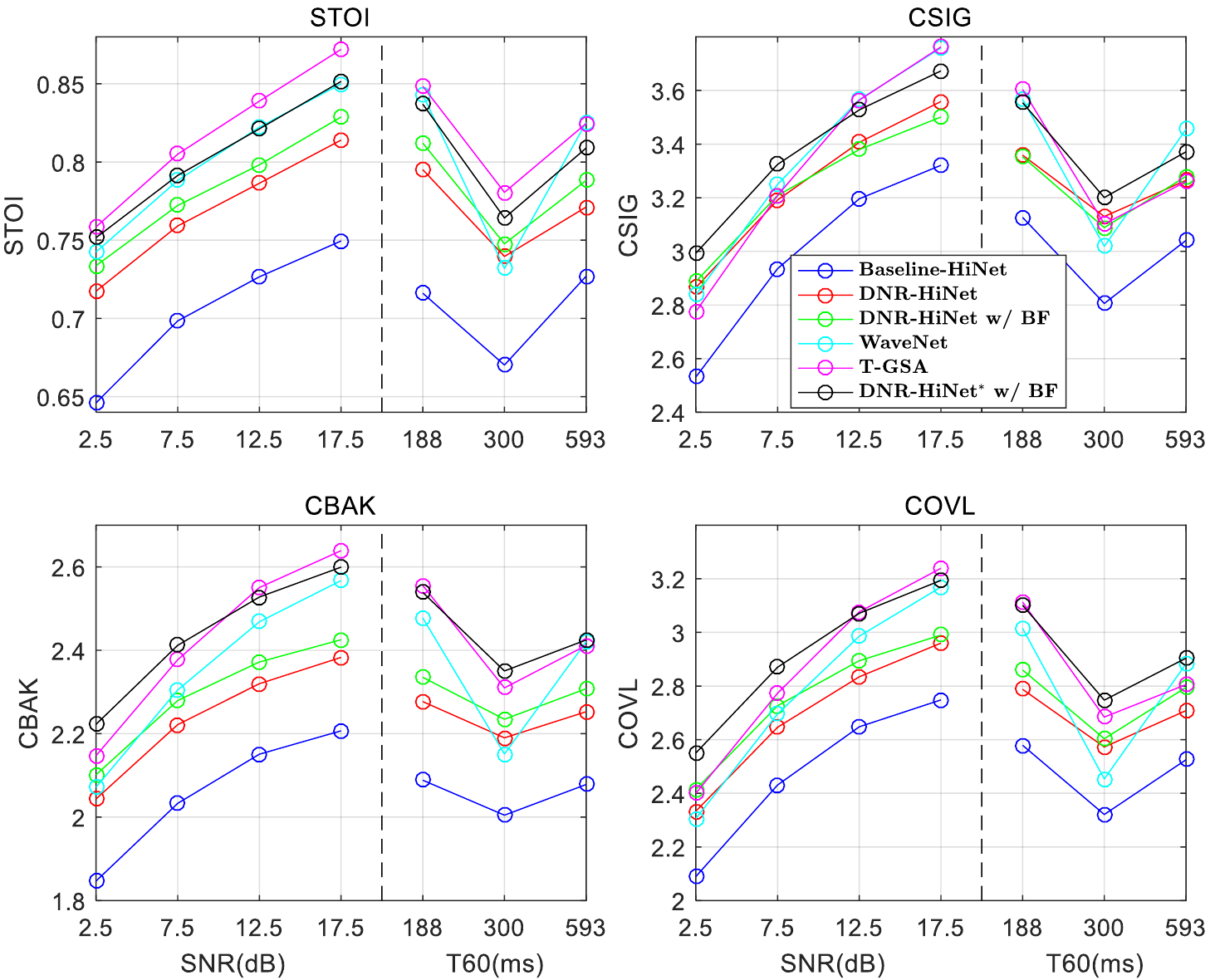}
    \caption{Objective evaluation results of different systems under different SNR and RIR conditions of test set.}
    \label{fig: objective}
    \vspace{-5mm}
\end{figure}

The average objective results on the test set are listed in Table \ref{tab_objective}.
Both \textbf{Baseline-NSF} and \textbf{Baseline-HiNet} outperformed their lower-bound models (i.e., \textbf{Baseline-NSF$'$} and \textbf{Baseline-HiNet$'$}), which indicate that both models can perform denoising and dereverberation to a limited degree if they are trained with the clean waveform as the target.
\textbf{Baseline-HiNet} had better noise suppression and overall performance than \textbf{Baseline-NSF} regarding CBAK and COVL.
However, the speech intelligibility of \textbf{Baseline-NSF} was better than that of \textbf{Baseline-HiNet} according to the STOI scores.

\textbf{DNR-HiNet} significantly outperformed \textbf{Baseline-HiNet} on all metrics, which indicates the effectiveness of the denoising and dereverberation structure in its DNR-ASP and the ineffectiveness of the baseline vocoders.
\textbf{DNR-HiNet w/ BF} achieved better scores than \textbf{DNR-HiNet} on all metrics except CSIG, but the difference in CSIG was very small.
This indicates that the extra BWE model in \textbf{DNR-HiNet w/ BF} generated higher quality clean LAS at a high frequency compared with \textbf{DNR-HiNet}, and the extra FRE model in \textbf{DNR-HiNet w/ BF} also improved the quality of the clean LAS by extending the frequency resolution.

Because the test set covered four SNRs and three RIRs, we also analyzed the objective results under different SNR and RIR conditions.
The results are shown in Figure \ref{fig: objective}.
As expected, the objective scores were higher when the SNR was larger.
In terms of dereverberation, all systems acquired lower scores when T60$=300ms$ than the other two cases, which may due to the fact that this RIR was selected from a different RIR dataset.
As Figure \ref{fig: objective} shows, the score differences among \textbf{Baseline-HiNet}, \textbf{DNR-HiNet}, and \textbf{DNR-HiNet w/ BF} were consistent across different SNR and RIR conditions and objective metrics.
Regarding CSIG, \textbf{DNR-HiNet} outperformed \textbf{Baseline-HiNet}, but there was little difference between \textbf{DNR-HiNet} and \textbf{DNR-HiNet w/ BF}. These conclusions are consistent with those given in Table \ref{tab_objective}.

\vspace{-1mm}
\subsubsection{Subjective evaluation}
\label{subsec: Subjective evaluation}
\vspace{-1mm}

We conducted two listening tests on the crowdsourcing platform Amazon Mechanical Turk\footnote{\url{https://www.mturk.com}}
with anti-cheating considerations \cite{buchholz2011crowdsourcing} to evaluate noise and reverberation suppression and speech quality, respectively.
In each test, 20 test utterances synthesized using each experimental model were evaluated by about 40 English native listeners.
\textbf{Baseline-NSF$'$} and \textbf{Baseline-HiNet$'$} were not included due to their significant poor objective results presented in Section \ref{subsec: Objective evaluation}.

\noindent
\textbf{Noise and reverberation suppression:}
The first test was to evaluate the suppression of noise and reverberation.
Listeners were asked to first listen to natural clean and natural noisy and reverberant audio tracks for reference.
They were then asked to listen to a few test audio tracks and assign a score from 1 to 9 to each audio, where a higher score denoted better suppression of noise and reverberation.

The results are listed in Table \ref{tab_subjective}.
Comparing \textbf{Baseline-NSF} with \textbf{Baseline-HiNet}, we can see that \textbf{Baseline-NSF} had better noise and reverberation suppression than \textbf{Baseline-HiNet} ($p$=0.035)\footnote{This is the $p$-value of a $t$-test. $p<0.05$ means that the difference between two compared models is significant.}.
As expected, the suppression scores of \textbf{DNR-HiNet} had higher means than those of \textbf{Baseline-HiNet} ($p\ll$0.05).
This suggests that the denoising and dereverberation structure in \textbf{DNR-HiNet} greatly helped to suppress noise and reverberation.
\textbf{DNR-HiNet w/ BF} further outperformed \textbf{DNR-HiNet} ($p$=0.027), which indicates the usefulness of the additional BWE and FRE models for noise and reverberation suppression.

\begin{table}
\vspace{-1mm}
\centering
    \caption{Subjective evaluation results of different systems. Here, the numbers in the table represent average scores with 95\% confidence interval. ``Group 1" and ``Group 2" represent the systems for comparison in Section \ref{subsec: Comparison among neural vocoders} and \ref{subsec: Challenge with SE methods} respectively.}
    \renewcommand\arraystretch{1.2}
    \resizebox{8.0cm}{2.2cm}{
    \begin{tabular}{c l | c | c}
        \hline
        \hline
         & Systems & Suppression score & MUSHRA score \\
         \hline
         \multirow{4}{*}{Group 1}
         & \textbf{Baseline-NSF} & 5.635$\pm$0.131 & 57.30$\pm$1.74 \\
         & \textbf{Baseline-HiNet} & 5.477$\pm$0.133 & 57.82$\pm$1.60 \\
         & \textbf{DNR-HiNet} & 5.774$\pm$0.128 & 60.51$\pm$1.60 \\
         & \textbf{DNR-HiNet w/ BF} & \textbf{5.939$\pm$0.128} & \textbf{61.73$\pm$1.55} \\
         \hline
         \multirow{6}{*}{Group 2}
         & \textbf{DNR-HiNet w/ BF} & \textbf{5.700$\pm$0.129} & \textbf{65.38$\pm$1.48} \\
         & \textbf{cIRM} & 4.975$\pm$0.138 & 55.27$\pm$1.88 \\
         & \textbf{SEGAN} & 4.873$\pm$0.155 & 49.06$\pm$2.07 \\
         & \textbf{WaveNet} & 5.396$\pm$0.130 & 62.18$\pm$1.59 \\
         & \textbf{T-GSA} & 5.624$\pm$0.121 & 62.28$\pm$1.58 \\
         & \textbf{DNR-HiNet$^*$ w/ BF} & \textbf{5.703$\pm$0.129} & \textbf{65.56$\pm$1.52} \\
        \hline
        \hline
    \end{tabular}}
\label{tab_subjective}
\vspace{-5mm}
\end{table}

\noindent
\textbf{Speech quality:}
The second test was a MUSHRA (MUltiple Stimuli with Hidden Reference and Anchor) test \cite{recommendation2001method} to evaluate the quality of generated speech of the different systems.
Listeners were asked to give a score between 0 and 100 to each test sample (the reference natural clean audio tracks had a maximum score of 100).

The results are listed in Table \ref{tab_subjective}.
While the two baselines \textbf{Baseline-HiNet} and \textbf{Baseline-NSF} achieve similar performance ($p$=0.46), \textbf{DNR-HiNet} had higher MUSHRA scores than the baselines ($p\ll$0.05), suggesting that the denoising and dereverberation structure in \textbf{DNR-HiNet} are also effective in improving the quality of synthetic speech.
\textbf{DNR-HiNet w/ BF} further outperformed \textbf{DNR-HiNet} in terms of the average MUSHRA score. Although the difference was not statistically significant ($p$=0.078), it indicates that the BWE and FRE models can assist the DNR-ASP in generating LAS with higher quality and improve the perceived quality of generated waveforms to some extent.

\vspace{-1mm}
\subsection{Comparison with SE methods}
\label{subsec: Challenge with SE methods}
\vspace{-1mm}

We then compared \textbf{DNR-HiNet w/ BF} with the SE methods mentioned in Section \ref{subsec: Experimental models}. For the SE task, the proposed model can directly use the phase of the input waveform rather than predicting it using the PSP, and such an SE-specific system is denoted as \textbf{DNR-HiNet$^*$ w/ BF}.

As shown in Table \ref{tab_objective}, while \textbf{DNR-HiNet w/ BF} outperformed \textbf{cIRM} and \textbf{SEGAN}\footnote{We also tried to train \textbf{cIRM} using a noisy database and got satisfactory results, which indicates that \textbf{cIRM} might not be suitable for dereverberation.},  it obtained lower scores than \textbf{WaveNet} and \textbf{T-GSA} for all objective metrics.
Interestingly, \textbf{DNR-HiNet$^*$ w/ BF} surpassed \textbf{WaveNet} and \textbf{T-GSA} for CSIG, CBAK and COVL.
This suggests that \textbf{DNR-HiNet w/ BF} was limited by inaccurate phase prediction.
However, the speech intelligibility of \textbf{DNR-HiNet$^*$ w/ BF} was still worse than \textbf{T-GSA} as the STOI scores in Table \ref{tab_objective} and Figure \ref{fig: objective} show.
We can also see from Figure \ref{fig: objective} that \textbf{DNR-HiNet w/ BF} and \textbf{DNR-HiNet$^*$ w/ BF} achieved similar or higher scores for CSIG, CBAK, and COVL than \textbf{T-GSA} under low SNR and high T60 conditions, which indicates that our proposed method is more robust in heavily polluted noise and reverberation environments.

As the subjective results in Table \ref{tab_subjective} indicate, \textbf{DNR-HiNet w/ BF} significantly outperformed \textbf{cIRM} and \textbf{SEGAN} on both noise and reverberation suppression and MUSHRA tests ($p\ll$0.05 for all scores).
Although \textbf{DNR-HiNet w/ BF} had worse objective results than \textbf{WaveNet} and \textbf{T-GSA} in Section \ref{subsec: Objective evaluation}, the subjective results are encouraging.
As shown in Table \ref{tab_subjective}, \textbf{DNR-HiNet w/ BF} suppressed noise and reverberation better than \textbf{WaveNet} ($p\ll$0.05) while being comparable to \textbf{T-GSA} ($p$=0.32).
The speech generated by \textbf{DNR-HiNet w/ BF} also had better perceived quality than that of \textbf{WaveNet} and \textbf{T-GSA} ($p\ll$0.05), as the MUSHRA scores in Table \ref{tab_subjective} show.
\textbf{DNR-HiNet w/ BF} and \textbf{DNR-HiNet$^*$ w/ BF} had similar subjective performance ($p$=0.97 for suppression score and $p$=0.79 for MUSHRA score), as shown in Table \ref{tab_subjective}.
This suggests that, although the predicted phase spectra in \textbf{DNR-HiNet w/ BF} degraded objective evaluation scores, they are not perceptually different from the natural phase extracted from the input noisy and reverberant waveform.
This also confirms that \textbf{DNR-HiNet w/ BF} predicted amplitude spectra with higher quality than \textbf{T-GSA}, suggesting the effectiveness of the proposed denoising and dereverberation structure in the DNR-ASP.

In summary, although \textbf{DNR-HiNet w/ BF} was originally TTS-oriented, it can be applied to an SE task (i.e., \textbf{DNR-HiNet$^*$ w/ BF}) where the model can directly use the phase spectra from the input noisy and reverberant speech.
For the SE task, the results indicate that \textbf{DNR-HiNet$^*$ w/ BF} is comparable to or better than \textbf{WaveNet} and \textbf{T-GSA} ($p$=0.26 for suppression score between \textbf{DNR-HiNet$^*$ w/ BF} and \textbf{T-GSA}, and $p\ll$0.05 for other scores) considering the subjective evaluations.
Therefore, our proposed method can be applied to both the vocoder and SE tasks with good performance.

\vspace{-1mm}
\section{Conclusion}
\label{sec: Conclusion}
\vspace{-1mm}

In this paper, we proposed the DNR-HiNet vocoder for the denoising and dereverberation task, with the goal of predicting a clean speech waveform from input noisy and reverberant acoustic features.
Compared with the original HiNet vocoder, we mainly modified the ASP as the DNR-ASP.
The DNR-ASP in the DNR-HiNet vocoder can  predict the clean LAS from input low-dimensional noisy and reverberant acoustic features.
First, the DNR-ASP adopts three different modules to predict the noisy and reverberant LAS, noise LAS, and RIR respectively, and performs initial denoising and dereverberation by deconvolution and noise subtraction in the frequency domain.
The initial enhanced results then pass through a neural network to obtain the final clean LAS.
We also introduced a BWE model and FRE model to the DNR-ASP to further improve the quality of the generated clean LAS.
Objective and subjective evaluation results indicated that the DNR-HiNet vocoder achieved better performance than the baseline HiNet and NSF vocoders on the denoising and dereverberation waveform generation task given degraded acoustic features.
Further experiments show that the proposed method can be also applied to SE task, and its performance was comparable with several advanced SE methods.
For future work, we plan to further improve the performance and reduce the model size of the DNR-HiNet vocoder.

\vspace{0.0mm}
\noindent\textbf{Acknowledgments:}
This work was partially supported by a JST CREST Grant (JPMJCR18A6, VoicePersonae project), Japan, MEXT KAKENHI Grants (19K24371, 16H06302, 17H04687, 18H04120, 18H04112, 18KT0051), Japan, and the National Natural Science Foundation of China under Grant 61871358 and the China Scholarship Council (CSC). The experiments were partially conducted on TSUBAME 3.0 of Tokyo Institute of Technology.

\vfill\pagebreak
\bibliographystyle{IEEEbib}
\bibliography{refs}

\begin{thebibliography}{10}

\bibitem{shen2018natural}
Jonathan Shen, Ruoming Pang, Ron~J Weiss, Mike Schuster, Navdeep Jaitly,
  Zongheng Yang, Zhifeng Chen, Yu~Zhang, Yuxuan Wang, Rj~Skerrv-Ryan, and
  Others,
\newblock ``{Natural {TTS} synthesis by conditioning {WaveNet} on {Mel}
  spectrogram predictions},''
\newblock in {\em Proc. ICASSP}, 2018, pp. 4779--4783.

\bibitem{liu2018wavenet}
Li-Juan Liu, Zhen-Hua Ling, Yuan Jiang, Ming Zhou, and Li-Rong Dai,
\newblock ``Wave{N}et vocoder with limited training data for voice
  conversion,''
\newblock in {\em Proc. Interspeech}, 2018, pp. 1983--1987.

\bibitem{tamamori2017speaker}
Akira Tamamori, Tomoki Hayashi, Kazuhiro Kobayashi, Kazuya Takeda, and Tomoki
  Toda,
\newblock ``Speaker-dependent {W}ave{N}et vocoder,''
\newblock in {\em Proc. Interspeech}, 2017, pp. 1118--1122.

\bibitem{ai2018samplernn}
Yang Ai, Hong-Chuan Wu, and Zhen-Hua Ling,
\newblock ``Sample{RNN}-based neural vocoder for statistical parametric speech
  synthesis,''
\newblock in {\em Proc. ICASSP}, 2018, pp. 5659--5663.

\bibitem{lorenzo2018robust}
Jaime Lorenzo-Trueba, Thomas Drugman, Javier Latorre, Thomas Merritt, Bartosz
  Putrycz, Roberto Barra-Chicote, Alexis Moinet, and Vatsal Aggarwal,
\newblock ``Towards achieving robust universal neural vocoding,''
\newblock in {\em Proc. Interspeech}, 2019, pp. 181--185.

\bibitem{oord2017parallel}
Aaron van~den Oord, Yazhe Li, Igor Babuschkin, Karen Simonyan, Oriol Vinyals,
  Koray Kavukcuoglu, George van~den Driessche, Edward Lockhart, Luis~C Cobo,
  Florian Stimberg, et~al.,
\newblock ``Parallel {W}ave{N}et: Fast high-fidelity speech synthesis,''
\newblock in {\em Proc. ICML}, 2018, pp. 3918--3926.

\bibitem{ping2018clarinet}
Wei Ping, Kainan Peng, and Jitong Chen,
\newblock ``Clari{N}et: Parallel wave generation in end-to-end
  text-to-speech,''
\newblock in {\em Proc. ICLR}, 2019.

\bibitem{prenger2018waveglow}
Ryan Prenger, Rafael Valle, and Bryan Catanzaro,
\newblock ``Wave{G}low: A flow-based generative network for speech synthesis,''
\newblock in {\em Proc. ICASSP}, 2019, pp. 3617--3621.

\bibitem{wangNSFall}
Xin Wang, Shinji Takaki, and Junichi Yamagishi,
\newblock ``Neural source-filter waveform models for statistical parametric
  speech synthesis,''
\newblock {\em IEEE/ACM Transactions on Audio, Speech, and Language
  Processing}, vol. 28, pp. 402--415, 2020.

\bibitem{ai2020neural}
Yang Ai and Zhen-Hua Ling,
\newblock ``{A neural vocoder with hierarchical generation of amplitude and
  phase spectra for statistical parametric speech synthesis},''
\newblock {\em IEEE/ACM Transactions on Audio, Speech, and Language
  Processing}, vol. 28, pp. 839--851, 2020.

\bibitem{ai2020knowledge}
Yang Ai and Zhen-Hua Ling,
\newblock ``Knowledge-and-data-driven amplitude spectrum prediction for
  hierarchical neural vocoders,''
\newblock in {\em Proc. Interspeech}, 2020.

\bibitem{valentini2018speech}
Cassia Valentini-Botinhao and Junichi Yamagishi,
\newblock ``{Speech enhancement of noisy and reverberant speech for
  text-to-speech},''
\newblock {\em IEEE/ACM Transactions on Audio, Speech, and Language
  Processing}, vol. 26, no. 8, pp. 1420--1433, 2018.

\bibitem{ai2019dnn}
Yang Ai, Jing-Xuan Zhang, Liang Chen, and Zhen-Hua Ling,
\newblock ``{DNN}-based spectral enhancement for neural waveform generators
  with low-bit quantization,''
\newblock in {\em Proc. ICASSP}, 2019, pp. 7025--7029.

\bibitem{xu2014regression}
Yong Xu, Jun Du, Li-Rong Dai, and Chin-Hui Lee,
\newblock ``A regression approach to speech enhancement based on deep neural
  networks,''
\newblock {\em IEEE/ACM Transactions on Audio, Speech, and Language
  Processing}, vol. 23, no. 1, pp. 7--19, 2014.

\bibitem{kim2020t}
Jaeyoung Kim, Mostafa El-Khamy, and Jungwon Lee,
\newblock ``{T-GSA}: Transformer with {G}aussian-weighted self-attention for
  speech enhancement,''
\newblock in {\em Proc. ICASSP}, 2020, pp. 6649--6653.

\bibitem{williamson2015complex}
Donald~S Williamson, Yuxuan Wang, and DeLiang Wang,
\newblock ``Complex ratio masking for monaural speech separation,''
\newblock {\em IEEE/ACM transactions on audio, speech, and language
  processing}, vol. 24, no. 3, pp. 483--492, 2015.

\bibitem{liu2019supervised}
Yun Liu, Hui Zhang, Xueliang Zhang, and Linju Yang,
\newblock ``Supervised speech enhancement with real spectrum approximation,''
\newblock in {\em Proc. ICASSP}, 2019, pp. 5746--5750.

\bibitem{su2019perceptually}
Jiaqi Su, Adam Finkelstein, and Zeyu Jin,
\newblock ``Perceptually-motivated environment-specific speech enhancement,''
\newblock in {\em Proc. ICASSP}, 2019, pp. 7015--7019.

\bibitem{rethage2018wavenet}
Dario Rethage, Jordi Pons, and Xavier Serra,
\newblock ``A {W}avenet for speech denoising,''
\newblock in {\em Proc. ICASSP}, 2018, pp. 5069--5073.

\bibitem{pascual2017segan}
Santiago Pascual, Antonio Bonafonte, and Joan Serr{\`a},
\newblock ``{SEGAN}: Speech enhancement generative adversarial network,''
\newblock {\em arXiv preprint arXiv:1703.09452}, 2017.

\bibitem{macartney2018improved}
Craig Macartney and Tillman Weyde,
\newblock ``Improved speech enhancement with the {W}ave-{U}-{N}et,''
\newblock {\em arXiv preprint arXiv:1811.11307}, 2018.

\bibitem{oord2016wavenet}
Aaron van~den Oord, Sander Dieleman, Heiga Zen, Karen Simonyan, Oriol Vinyals,
  Alex Graves, Nal Kalchbrenner, Andrew Senior, and Koray Kavukcuoglu,
\newblock ``Wave{N}et: A generative model for raw audio,''
\newblock in {\em 9th ISCA Speech Synthesis Workshop}, 2016, pp. 125--125.

\bibitem{ai2020reverberation}
Yang Ai, Xin Wang, Junichi Yamagishi, and Zhen-Hua Ling,
\newblock ``Reverberation modeling for source-filter-based neural vocoder,''
\newblock in {\em Proc. Interspeech}, 2020.

\bibitem{li2020noise}
Haoyu Li and Junichi Yamagishi,
\newblock ``Noise tokens: Learning neural noise templates for environment-aware
  speech enhancement,''
\newblock in {\em Proc. Interspeech}, 2020.

\bibitem{vaswani2017attention}
Ashish Vaswani, Noam Shazeer, Niki Parmar, Jakob Uszkoreit, Llion Jones,
  Aidan~N Gomez, {\L}ukasz Kaiser, and Illia Polosukhin,
\newblock ``Attention is all you need,''
\newblock in {\em Advances in neural information processing systems}, 2017, pp.
  5998--6008.

\bibitem{gulrajani2017improved}
Ishaan Gulrajani, Faruk Ahmed, Martin Arjovsky, Vincent Dumoulin, and Aaron~C
  Courville,
\newblock ``Improved training of {W}asserstein {GAN}s,''
\newblock in {\em Advances in neural information processing systems}, 2017, pp.
  5767--5777.

\bibitem{mescheder2018training}
Lars Mescheder, Andreas Geiger, and Sebastian Nowozin,
\newblock ``Which training methods for {GAN}s do actually converge?,''
\newblock {\em arXiv preprint arXiv:1801.04406}, 2018.

\bibitem{jeub2009binaural}
Marco Jeub, Magnus Schafer, and Peter Vary,
\newblock ``A binaural room impulse response database for the evaluation of
  dereverberation algorithms,''
\newblock in {\em Proc. ICDSP}, 2009, pp. 1--5.

\bibitem{hadad2014multichannel}
Elior Hadad, Florian Heese, Peter Vary, and Sharon Gannot,
\newblock ``Multichannel audio database in various acoustic environments,''
\newblock in {\em Proc. IWAENC}, 2014, pp. 313--317.

\bibitem{wen2006evaluation}
Jimi~YC Wen, Nikolay~D Gaubitch, Emanuel~AP Habets, Tony Myatt, and Patrick~A
  Naylor,
\newblock ``Evaluation of speech dereverberation algorithms using the mardy
  database,''
\newblock in {\em Proc. IWAENC}, 2006, pp. 1--4.

\bibitem{kasi2002yet}
Kavita Kasi and Stephen~A Zahorian,
\newblock ``Yet another algorithm for pitch tracking,''
\newblock in {\em Proc. ICASSP}, 2002, vol.~1, pp. 361--364.

\bibitem{wang2019}
Xin Wang and Junichi Yamagishi,
\newblock ``Neural harmonic-plus-noise waveform model with trainable maximum
  voice frequency for text-to-speech synthesis,''
\newblock in {\em Proc. SSW}, ISCA, sep 2019, pp. 1--6, ISCA.

\bibitem{taal2010short}
Cees~H Taal, Richard~C Hendriks, Richard Heusdens, and Jesper Jensen,
\newblock ``A short-time objective intelligibility measure for time-frequency
  weighted noisy speech,''
\newblock in {\em Proc. ICASSP}, 2010, pp. 4214--4217.

\bibitem{hu2006evaluation}
Yi~Hu and Philipos~C Loizou,
\newblock ``Evaluation of objective measures for speech enhancement,''
\newblock {\em IEEE/ACM Transactions on Audio, Speech, and Language
  Processing}, vol. 16, no. 1, pp. 229--238, 2008.

\bibitem{buchholz2011crowdsourcing}
Sabine Buchholz and Javier Latorre,
\newblock ``Crowdsourcing preference tests, and how to detect cheating,''
\newblock in {\em Proc. Interspeech}, 2011, pp. 3053--3056.

\bibitem{recommendation2001method}
ITUR Recommendation,
\newblock ``Method for the subjective assessment of intermediate sound quality
  ({MUSHRA}),''
\newblock {\em ITU, BS}, pp. 1543--1, 2001.

\end{thebibliography}

\end{document}